\documentclass[aps,prd,preprint,groupedaddress %,twocolumn,10pt
]{revtex4-1}
\usepackage[mathscr]{euscript}
\usepackage{float,epsfig}
\usepackage{bm}% bold math
\usepackage{graphicx}% Include figure files
\usepackage{subfigure}
\usepackage[colorlinks=true,linkcolor=blue,urlcolor=blue]{hyperref}
\newcommand{\bea}{\begin{eqnarray}}
\newcommand{\eea}{\end{eqnarray}}
\newcommand{\beq}{\begin{equation}}
\newcommand{\eeq}{\end{equation}}

\def\/{\over}

\begin{document}

% Use the \preprint command to place your local institutional report
% number in the upper righthand corner of the title page in preprint mode.
% Multiple \preprint commands are allowed.
% Use the 'preprintnumbers' class option to override journal defaults
% to display numbers if necessary
%\preprint{}

%Title of paper
\title{Quantum gravitational interaction between two objects induced by external gravitational radiation fields}

% repeat the \author .. \affiliation  etc. as needed
% \email, \thanks, \homepage, \altaffiliation all apply to the current
% author. Explanatory text should go in the []'s, actual e-mail
% address or url should go in the {}'s for \email and \homepage.
% Please use the appropriate macro foreach each type of information

% \affiliation command applies to all authors since the last
% \affiliation command. The \affiliation command should follow the
% other information
% \affiliation can be followed by \email, \homepage, \thanks as well.
\author{Yongshun Hu, Jiawei Hu\footnote{jwhu@hunnu.edu.cn}, Hongwei Yu\footnote{hwyu@hunnu.edu.cn} 
}
%\email[]{Your e-mail address}
%\homepage[]{Your web page}
%\thanks{}
%\altaffiliation{}
\affiliation{$^{1}$Department of Physics and Synergetic Innovation Center for Quantum Effects and Applications, Hunan Normal University, Changsha, Hunan 410081, China %\\
}

%Collaboration name if desired (requires use of superscriptaddress
%option in \documentclass). \noaffiliation is required (may also be
%used with the \author command).
%\collaboration can be followed by \email, \homepage, \thanks as well.
%\collaboration{}
%\noaffiliation

%\date{\today}

\begin{abstract}

We explore, in the framework of linearized quantum gravity,  the induced gravitational interaction between two gravitationally polarizable objects in their ground states in the presence  of an external quantized gravitational radiation field. The interaction energy  decreases as $r^{-5}$ in the near regime, and oscillates with a decreasing amplitude proportional to $r^{-1}$ in the far regime, where $r$ is the distance between the two objects. The interaction can be either attractive or repulsive  depending on the propagation direction, polarization and frequency of the external gravitational field. That is, the induced interaction can be manipulated by varying the relative direction between the orientation of the objects with respect to the propagation direction of the incident gravitational radiation.

\end{abstract}

% insert suggested PACS numbers in braces on next line
% insert suggested keywords - APS authors don't need to do this
%\keywords{}

%\maketitle must follow title, authors, abstract, \pacs, and \keywords
\maketitle

% body of paper here - Use proper section commands
% References should be done using the \cite, \ref, and \label commands
\section{Introduction}
\label{sec_in}
\setcounter{equation}{0}
%%%%%%%%%%%%%%%%%%%%%%%%%%%%%%%%%%%%%%%%%%%%%%%%%%
It is well known that in a quantum sense, there inevitably exist quantum vacuum fluctuations, which may induce some novel effects. One of the most famous examples is the electromagnetic Casimir-Polder (CP)  interaction \cite{CP}.
In general, fluctuating electromagnetic fields in vacuum induce instantaneous electric dipole moments in neutral atoms,  which then couple with each other via the exchange of virtual photons to yield an interaction energy. For atoms or molecules in different states, such CP interactions  behave differently in terms of  distance-dependence  \cite{PT,Salam,Power1993pra,McLone1965,Gomberoff1966,Power1993Chem,Power1995,Rizzuto2004, Sherkunov2007, Preto2013,Donaire2015,Milonni2015,Berman2015,Jentschura2017}.  For example, the interatomic or intermolecular interaction behaves as $r^{-6}$ and $r^{-7}$ in the near and far regimes respectively when the atoms or molecules are in their ground states \cite{CP}, while it behaves as $r^{-3}$ and $r^{-1}$ in the near and far regimes respectively when they are prepared in  a symmetric/antisymmetric  entangled state \cite{PT}.

Likewise, one may also expect a gravitational CP-like  interaction if one accepts that basic quantum principles are also applicable to gravity. Unfortunately, a full theory of quantum gravity is elusive at present. Even though, one may still study  quantum gravitational effects at low energies in the framework of linearized quantum gravity \cite{yu1999,Ford1995},  the basic idea of which is to
express the spacetime metric as a sum of the flat background spacetime metric and  a linearized perturbation,
%consider a flat background spacetime with a linearized perturbation propagating upon it,
and quantize the perturbation part in the canonical approach. Based on linearized quantum gravity,
%One such typical example is the quantum light-cone fluctuations \cite{yu1999,Ford1995,yu2009}.
the gravitational CP-like  interactions between two gravitationally polarizable objects in their ground states, and between one gravitationally polarizable object and a gravitational boundary, have  recently been studied in Refs. \cite{Ford2016,Wu2016,Wu2017,Holstein2017,Hu2017,yu2018}.
%In a general way, vacuum fluctuations of gravitational fields induce instantaneous gravitational quadrupole moments in gravitationally polarizable objects with which then interact each other via the exchange of virtual gravitons to obtain an interaction energy.
Similar to the electromagnetic case, the behaviors of gravitational CP-like  interactions are  significantly different when the gravitationally polarizable objects are prepared in different states. For example, the gravitational CP-like  potential is found to be proportional to $r^{-10}$ and $r^{-11}$ in the near and far regimes respectively when the two objects are in their ground states \cite{Ford2016,Wu2016,Wu2017,Holstein2017}, while it behaves as $r^{-5}$ and $r^{-1}$ in the near and far regimes respectively when the two objects are in a symmetric/antisymmetric  entangled state \cite{Hu2019}.
%It shows that the inter-object interaction has been significantly modified by changing the states of the gravitationally polarizable objects.
%{\bf It shows that the CP-like gravitational interactions can be changed in the way similar to the electromagnetic cases, i.e. the interaction potential can be changed by changing the states of the objects.}

Naturally, a question arises as to whether such quantum gravitational effects can be modified or enhanced in certain circumstances. Fortunately, there are similar examples in quantum electrodynamics. For example, the interaction between two ground-state atoms or molecules is found to be modified in the presence of external electromagnetic radiation fields \cite{Thirunamachandran1980,Milonni1992,Milonni1996,Salam2006,Salam2007}. That is, the externally applied electromagnetic field induces dipole moments in atoms or molecules, which are coupled with each other via the exchange of a single virtual photon, and an  interaction is induced. This process is clearly different from the case without external electromagnetic fields, which arises from  two-photon exchange. Similarly, in the gravitational case, one may expect that the quantum gravitational quadrupole-quadrupole interactions will also be modified in the presence of an external gravitational radiation field.

In this paper, we explore the quantum gravitational quadrupole-quadrupole interaction between a pair of gravitationally polarizable objects in their ground states, which are subjected to a weak external gravitational radiation field based on the leading-order perturbation theory in the framework of linearized quantum gravity. First, we describe in  details  the system we deal with. Then, we obtain the general expression for the interaction energy  between the two objects. Finally, we discuss our results in specific cases and obtain the corresponding interaction potentials. Throughout this paper, the Einstein summation convention for repeated indices is assumed, and the Latin indices run from $1$ to $3$ while the Greek indices run from $0$ to $3$. Units with $\hbar=c=16\pi G=1$ are applied, where $\hbar$ is the reduced Planck constant, $c$ is the speed of light and  $G$ is  the Newtonian gravitational constant.

%%%%%%%%%%%%%%%%%%%%%%%%%%%%%%%%%%%%%%%%%%%%%%%%%%
\section{Basic equations}
\label{sec_ba}
%\setcounter{equation}{0}
%%%%%%%%%%%%%%%%%%%%%%%%%%%%%%%%%%%%%%%%%%%%%%%%%%
We consider two gravitationally polarizable objects (labeled as A and B) coupled with a bath of fluctuating gravitational fields in vacuum, which are subjected to a weak external gravitational radiation field. The objects A and B are modeled as two-level systems with two internal energy levels, $\pm \frac{1}{2}\omega_0$, associated with the eigenstates $|g\rangle$ and $|e\rangle$, respectively. The total Hamiltonian is
\beq \label{Hamiltonian}
H=H_F+H_R+H_S+H_I,
\eeq
where $H_{F}$ is the Hamiltonian of the fluctuating vacuum gravitational field, $H_R$  the Hamiltonian of the  external gravitational radiation field, $H_{S}$  the Hamiltonian of the two-level systems (A and B), and $H_I$ the interaction Hamiltonian between the objects and the gravitational fields. Here $H_I$ takes the form
\beq\label{HI}
H_I=-\frac{1}{2}Q^{A}_{ij}[\epsilon_{ij}(\vec x_A)+E_{ij}(\vec x_A)]-\frac{1}{2}Q^{B}_{ij}[\epsilon_{ij}(\vec x_B)+E_{ij}(\vec x_{B})],
\eeq
where $Q^{A(B)}_{ij}$ is the induced quadrupole moment of the object A (B), $\epsilon_{ij}$ is the gravitoelectric tensor characterizing the weak external gravitational radiation field, and $E_{ij}$ is the gravitoelectric tensor of the fluctuating vacuum gravitational fields defined as $E_{ij}=C_{0i0j}$ by an analogy between the linearized Einstein field equations and the Maxwell equations~\cite{WB}, where $C_{\mu\nu\alpha\beta}$ is the Weyl tensor. We write the spacetime metric  $g_{\mu\nu}$ as a sum of the flat spacetime metric $\eta_{\mu\nu}$  and a linearized perturbation  $h_{\mu\nu}$,  then  the gravitoelectric tensor $E_{ij}$ can be expressed as (in the transverse traceless gauge)
\beq\label{Eij}
E_{ij}=\frac{1}{2}\ddot h_{ij}.
\eeq
Suppose that the linearized perturbation $h_{\mu\nu}$ is quantized, in this regard, we can decompose $h_{\mu\nu}$ into positive and negative frequency parts $h_{\mu\nu}^+$ and $h_{\mu\nu}^-$, respectively, and define  the gravitational vacuum state $|0\rangle$ as
\beq
h_{\mu\nu}^+ |0\rangle=0, \quad \langle 0|h_{\mu\nu}^-=0.
\eeq
It follows immediately that $\langle 0|h_{\mu\nu}|0\rangle =0$. In general, however, $\langle 0|(h_{\mu\nu})^2|0\rangle \neq0$, where the expectation value is understood to be suitably renormalized. In the transverse traceless gauge, the quantized gravitational perturbations have only spatial components $h_{i j}$, which takes the standard form
%Here $h_{ij}$ describes the quantized fluctuating vacuum gravitational fields,  which takes the standard form
\beq\label{hij}
h_{ij}=\sum_{\vec p,\lambda} \sqrt{\frac{1}{2\omega(2\pi)^3}} \left[a_{\lambda}(\vec p) e^{(\lambda)}_{ij} e^{i(\vec p\cdot \vec x-\omega t)}+\text{H.c.}\right],
\eeq
where $a_{\lambda}(\vec p)$ is the annihilation operator of the gravitational vacuum field with wave vector $\vec p$ and polarization $\lambda$, $e^{(\lambda)}_{ij}$ are polarization tensors, $\omega=|\vec p|=(p_{x}^2+p_{y}^2+p_{z}^2)^{1/2}$, and H.c. denotes the Hermitian conjugate.
As for the weak external gravitational radiation field, we assume that it can be described as a quantized monochromatic  gravitational wave containing $N$ gravitons. Then, the corresponding gravitoelectric tensor $\epsilon_{ij}$ can be  given as
\beq\label{epsilon ij}
\epsilon_{ij}=-\frac{1}{2} \sqrt{\frac{\omega_R^3 \rho_n}{2N(2\pi)^3}} \left[b(\vec k)e_{ij}^{(\varepsilon)} e^{i(\vec k\cdot \vec x-\omega_R t)}+H.c.\right],
\eeq
where $\rho_n$ is the number density of gravitons, $b(\vec k)$ and $e_{ij}^{(\varepsilon)}$ are respectively the corresponding annihilation operator and the polarization tensors with $|\vec k|=\omega_{R}$, and $\varepsilon$ labels the polarization state.

In the absence of an external gravitational field, the interaction between a pair of ground-state  objects coupled with a bath of fluctuating gravitational fields in vacuum is a fourth-order effect \cite{Ford2016,Wu2016,Wu2017}: The gravitational vacuum fluctuations induce quadrupole moments in the two objects, which are correlated  and   an interaction energy is thus induced. Physically speaking, such an induced interaction originates from  vacuum fluctuations and arises through the exchange of a pair of virtual gravitons between the two objects. In the present case, the leading interaction between quadrupole moments induced by the external gravitational radiation field will also be a fourth-order effect.
%in coupling with the gravitational vacuum fluctuations and the external gravitational field.
However, the difference is that  the quadrupole moments are now induced by the external gravitational field, which are then correlated to each other through gravitational vacuum fluctuations. That is, a real graviton will be scattered by a pair of objects which are coupled  via the exchange of a virtual graviton, and an interaction  is then induced, which is analogous to the electromagnetic case \cite{Thirunamachandran1980}.

\begin{figure}[htbp]
  \centering
  % Requires \usepackage{graphicx}
  \includegraphics[width=0.3\textwidth]{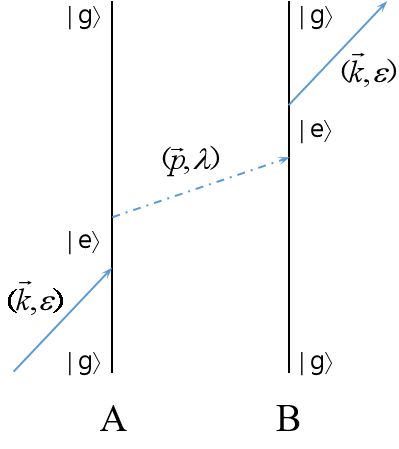}\\
  \caption{A typical time-ordered diagram for the calculation of inter-object interaction in the existence of an external quantized gravitational field. The blue solid line represents a real graviton, while the dotted one represents a virtual one. }\label{T1}
\end{figure}

\begin{figure}[htbp]
  \centering
  % Requires \usepackage{graphicx}
  \includegraphics[width=0.5\textwidth]{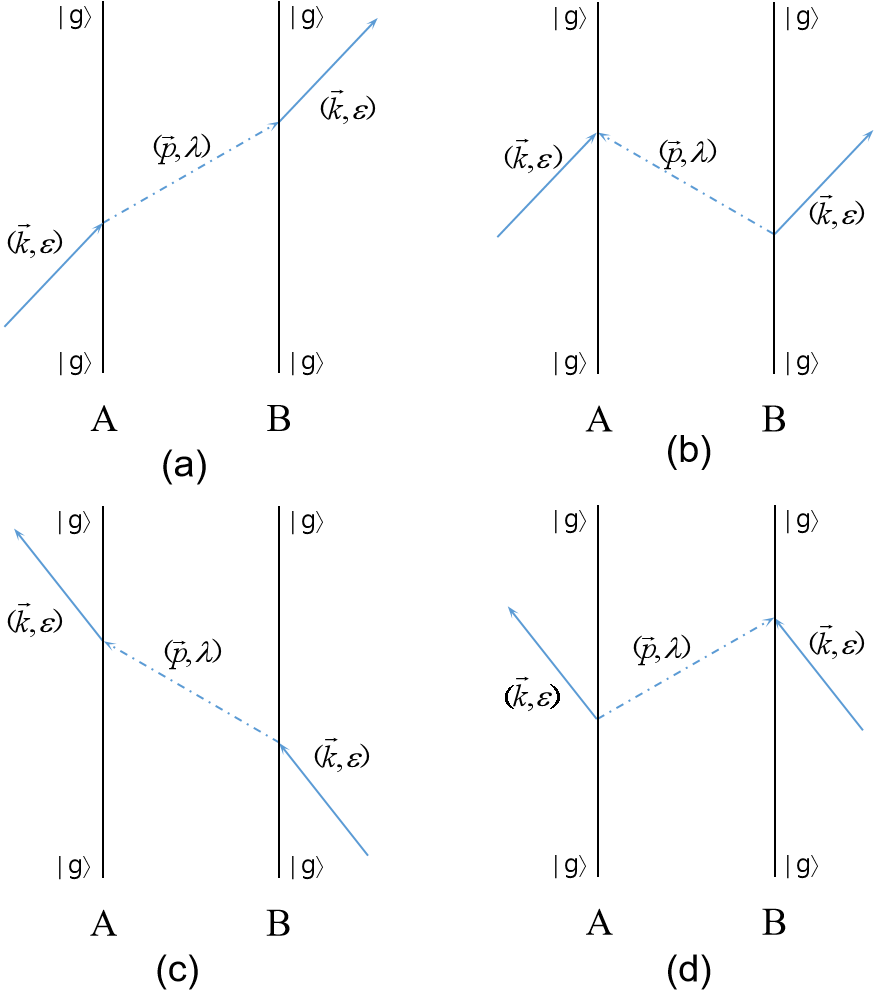}\\
  \caption{Four time-ordered diagrams represent the four contributing terms in the second order perturbation theory.}\label{4F}
\end{figure}

We choose the initial state of the system to be
\beq\label{phi}
|\phi\rangle=|g_A g_B\rangle|0\rangle|N\rangle,
\eeq
where $|g_A g_B\rangle$ is the ground state of the objects, $|0\rangle$ is the vacuum state of the fluctuating gravitational  field, and $|N\rangle$ is the number state of the external gravitational radiation field.
The initial energy of the whole system is $E_{\phi}=E_0+N \omega_R$, where $E_0$ denotes the ground-state energy of the objects and fluctuating gravitational field in vacuum. The leading contribution to the interaction energy  can be obtained from  fourth-order perturbation theory, which contains $48$ possible Feynman diagrams in our case, and a typical one is shown in Fig.~\ref{T1}.
However, the calculations can be greatly simplified by collapsing the two one-graviton interaction vertices in the time-ordered diagrams, which can be described as an effective two-graviton interaction Hamiltonian. To do this, we introduce the gravitational  polarizability of the objects, and, for simplicity, assume that  the objects are isotropically polarizable. Then, the induced quadrupole can be expressed as
\beq\label{Qij}
Q^{A(B)}_{ij}=\alpha^{(\varepsilon)}_{A(B)} \epsilon_{ij},
\eeq
where $\alpha^{(\varepsilon)}_{A(B)}$ is the isotropic polarizability of object A(B). In order to calculate the interaction, we only keep the corresponding terms after substituting Eq. (\ref{Qij}) into Eq. (\ref{HI}). Then, the effective Hamiltonian takes the form
\beq\label{Heff}
H^{eff}_I=-\frac{1}{2} \alpha^{(\varepsilon)}_A \epsilon_{ij}(\vec x_A) E_{ij}(\vec x_A)-\frac{1}{2} \alpha^{(\varepsilon)}_B \epsilon_{ij}(\vec x_B) E_{ij}(\vec x_B).
\eeq
The interaction energy  can  be calculated based on the second order perturbation theory
\beq
\Delta E=-\sum_{I}\frac{\langle \phi|H^{eff}_{I}|I\rangle\langle I|H^{eff}_{I}|\phi\rangle}{E_{I}-E_{\phi}},
\eeq
with only four contributing time-ordered diagrams as shown  in Fig. \ref{4F}.
%For these four diagrams, the  intermediate state $|I\rangle=|g_A g_B\rangle|1_{\vec p \lambda}\rangle|N-1\rangle$ correspond to $(a)(c)$, while $|I\rangle=|g_A g_B\rangle|1_{\vec p \lambda}\rangle|N+1\rangle$ correspond to $(b)(d)$.
Summing up all the contributions, the interaction energy can be expressed as
\bea\label{Eab}
\nonumber \Delta E_{AB}&=&-\frac{\omega^3_R \rho_n}{256(2\pi)^6}\alpha^{(\varepsilon)}_{A}\alpha^{(\varepsilon)}_{B} e^{(\varepsilon)}_{ij}e^{(\varepsilon)}_{kl}\cos{(\vec k \cdot \vec r)}\int d^3\vec p \sum_{\lambda}e^{(\lambda)}_{ij}e^{(\lambda)}_{kl} \frac{\omega^3 }{\omega-\omega_R}e^{i\vec p\cdot\vec r} \\
&-&\frac{(N+1)\omega^3_R \rho_n}{256N(2\pi)^6}\alpha^{(\varepsilon)}_{A}\alpha^{(\varepsilon)}_{B} e^{(\varepsilon)}_{ij}e^{(\varepsilon)}_{kl}\cos{(\vec k \cdot \vec r)}\int d^3\vec p \sum_{\lambda}e^{(\lambda)}_{ij}e^{(\lambda)}_{kl} \frac{\omega^3}{\omega+\omega_R} e^{i\vec p\cdot\vec r},
\eea
where $\vec r= \vec x_A - \vec x_B$. Here the summation of polarization tensors in the transverse traceless gauge gives~\cite{yu1999}
\beq
\nonumber\sum_{\lambda}e^{(\lambda)}_{ij}e^{(\lambda)}_{kl}=\delta_{ik}\delta_{j l}+\delta_{il}\delta_{j k}-\delta_{ij}\delta_{k l}-\frac{1}{\omega^2}H_{ijkl}+\frac{1}{\omega^4}P_{ijkl},
\eeq
where
\beq
H_{ijkl}=\partial_i\partial_j\delta_{kl}+\partial_k\partial_l\delta_{ij}-\partial_i\partial_k\delta_{jl} -\partial_i\partial_l\delta_{jk}-\partial_j\partial_k\delta_{il}-\partial_j\partial_l\delta_{ik},\ P_{ijkl}=\partial_i\partial_j\partial_k\partial_l.
\eeq
For convenience, we define a gravitational radiation intensity $I_R$ in analogy to the electromagnetic case \cite{Buhmann2019}  as
\beq
I_R=\langle N|\epsilon^2_{ij}|N\rangle=\frac{\omega^3_R \rho_n}{8N(2\pi)^3}(2N+1),
\eeq
since the intensity of the radiation field should be proportional to the number of gravitons. For a large graviton number $N\gg1$, we have
\beq
I_R\simeq\frac{\omega^3_R \rho_n}{4(2\pi)^3}.
\eeq
Thus, the interaction energy  (\ref{Eab}) can be expressed as
\beq\label{fullEab}
\Delta E_{AB}=-\frac{I_R}{128\pi}\alpha^{(\varepsilon)}_{A}\alpha^{(\varepsilon)}_{B} e^{(\varepsilon)}_{ij}e^{(\varepsilon)}_{kl} \cos{(\vec k \cdot \vec r)} V_{ijkl},
\eeq
where
\beq
V_{ijkl}=\left[(\delta_{ik}\delta_{j l}+\delta_{il}\delta_{j k}-\delta_{ij}\delta_{k l})\omega^4_R -\omega^2_R H_{ijkl} +P_{ijkl}\right]\frac{\cos{\omega_R r}}{r}.
\eeq
After some algebraic manipulations, the full form of $V_{ijkl}$ is given by
\bea
\nonumber V_{ijkl}=\frac{1}{r^5}\bigg[&&(\delta_{ik}\delta_{jl}+\delta_{il}\delta_{jk}-\delta_{ij}\delta_{kl}+\hat r_i \hat r_j\delta_{kl}+\hat r_k \hat r_l\delta_{ij}-\hat r_i \hat r_k\delta_{jl}\\
\nonumber &&-\hat r_i \hat r_l\delta_{jk}-\hat r_j \hat r_k\delta_{il}-\hat r_j \hat r_l\delta_{ik}+\hat r_i \hat r_j\hat r_k \hat r_l)r^4\omega^4_R \cos{\omega_R r}\\
\nonumber &&+2(-\delta_{ik}\delta_{jl}-\delta_{il}\delta_{jk}+\delta_{ij}\delta_{kl}-\hat r_i \hat r_j\delta_{kl}-\hat r_k \hat r_l\delta_{ij}+2\hat r_j \hat r_k\delta_{il}\\
\nonumber &&+2\hat r_j \hat r_l\delta_{ik}+2\hat r_i \hat r_k\delta_{jl}+2\hat r_i \hat r_l\delta_{jk}-5\hat r_i \hat r_j\hat r_k \hat r_l)r^3\omega^3_R \sin{\omega_R r}\\
\nonumber &&+(-3\delta_{ik}\delta_{jl}-3\delta_{il}\delta_{jk}+\delta_{ij}\delta_{kl}+3\hat r_i \hat r_j\delta_{kl}+3\hat r_k \hat r_l\delta_{ij}+9\hat r_j \hat r_k\delta_{il}\\
\nonumber &&+9\hat r_j \hat r_l\delta_{ik}+9\hat r_i \hat r_k\delta_{jl}+9\hat r_i \hat r_l\delta_{jk}-45\hat r_i \hat r_j\hat r_k \hat r_l) r^2\omega^2_R\cos{\omega_R r}\\
\nonumber &&+3(\delta_{ik}\delta_{jl}+\delta_{il}\delta_{jk}+\delta_{ij}\delta_{kl}-5\hat r_i \hat r_j\delta_{kl}-5\hat r_k \hat r_l\delta_{ij}-5\hat r_j \hat r_k\delta_{il}\\
\nonumber &&-5\hat r_j \hat r_l\delta_{ik}-5\hat r_i \hat r_k\delta_{jl}-5\hat r_i \hat r_l\delta_{jk}+35\hat r_i \hat r_j\hat r_k \hat r_l)r\omega_R \sin{\omega_R r}\\
\nonumber &&+3(\delta_{ik}\delta_{jl}+\delta_{il}\delta_{jk}+\delta_{ij}\delta_{kl}-5\hat r_i \hat r_j\delta_{kl}-5\hat r_k \hat r_l\delta_{ij}-5\hat r_j \hat r_k\delta_{il}\\
&&-5\hat r_j \hat r_l\delta_{ik}-5\hat r_i \hat r_k\delta_{jl}-5\hat r_i \hat r_l\delta_{jk}+35\hat r_i \hat r_j\hat r_k \hat r_l)\cos{\omega_R r}\bigg],
\eea
where $\hat r_i$ is a component of the unit vector $\vec r /r$. The above result shows that the total interaction energy depends on the polarization, frequency and  propagation direction of the external gravitational radiation field. In the following, we consider two explicit examples.  %of different propagations and orientations.

First, when the propagation direction of the external gravitational radiation field is parallel to the orientation of the two objects,
%i.e.
or equivalently, the polarization plane is perpendicular to $\vec r$ , i.e., $\vec k \cdot \vec r=\omega_R r$  and $e^{(\varepsilon)}_{ij}\hat r_i=0$, Eq. (\ref{fullEab}) can be rewritten as
\bea\label{Eab para}
\nonumber\Delta E_{AB}=-\frac{I_R}{64\pi r^5}\alpha^{(\varepsilon)}_{A}\alpha^{(\varepsilon)}_{B} e^{(\varepsilon)}_{ij}e^{(\varepsilon)}_{ij} &&\Big(r^4\omega^4_R\cos^2{\omega_R r}-2r^3\omega^3_R\sin{\omega_R r}\cos{\omega_R r} -3r^2\omega^2_R\cos^2{\omega_R r}\\
&&+3r\omega_R\sin{\omega_R r}\cos{\omega_R r}+3\cos^2{\omega_R r}\Big),
\eea
where $e^{(\varepsilon)}_{ii}=0$ and $e^{(\varepsilon)}_{ij}=e^{(\varepsilon)}_{ji}$ have been applied. In the near regime, i.e., $\omega_R r\ll 1$, the leading term takes the form
\beq\label{19}
\Delta E_{AB}\simeq-\frac{3I_R}{64\pi r^5}\alpha^{(\varepsilon)}_{A}\alpha^{(\varepsilon)}_{B} e^{(\varepsilon)}_{ij}e^{(\varepsilon)}_{ij},
\eeq
while in the far regime, i.e., $\omega_R r\gg 1$,  it becomes
\beq\label{20}
\Delta E_{AB}
%&\simeq&-\frac{\omega^3_R I_R}{64\pi r^2}\alpha^{(\varepsilon)}_{A}\alpha^{(\varepsilon)}_{B} e^{(\varepsilon)}_{ij}e^{(\varepsilon)}_{ij}\sqrt{4+\omega_R^2 r^2}\cos{(\omega_R r-\phi)}\cos{\omega_R r}\\
\simeq-\frac{\omega^4_R I_R}{64\pi r}\alpha^{(\varepsilon)}_{A}\alpha^{(\varepsilon)}_{B} e^{(\varepsilon)}_{ij}e^{(\varepsilon)}_{ij}\cos{(\omega_R r+\phi)}\cos{\omega_R r},
\eeq
where $\phi=\arcsin{\frac{2}{\sqrt{4+\omega_R^2 r^2}}}$.   This shows that the gravitational interaction between two objects in the  presence of an external gravitational field decreases as $r^{-5}$ in the near regime, while in the far regime it oscillates with a decreasing amplitude proportional to $r^{-1}$.  Moreover, from Eqs. (\ref{19})-(\ref{20}), we observe that in the near regime, the interaction is always attractive, while in the far regime, it can be  attractive or repulsive depending on the frequency of the external gravitational field and the interobject distance.

%When $r=(n+\frac{1}{2})\pi/\omega_R$, the interaction vanishes, where $n$ is an integer. This can be understood by rewriting the interaction energy  Eq. (\ref{fullEab}) as
%\beq
%\Delta E_{AB}=-\frac{\alpha^{(\varepsilon)}_{A}\alpha^{(\varepsilon)}_{B}}{128\pi}\langle N|[\epsilon_{ij}(\vec x_A)\epsilon_{kl}(\vec x_B) +\epsilon_{kl}(\vec x_B) \epsilon_{ij}(\vec x_A)]|N\rangle V_{ijkl},
%\eeq
%where
%\beq
%\langle N|\epsilon_{ij}(\vec x_A)\epsilon_{kl}(\vec x_B) +\epsilon_{kl}(\vec x_B) \epsilon_{ij}(\vec x_A) |N\rangle=\frac{(2N+1)\rho_n}{8N(2\pi)^3}\omega^3_R e^{(\varepsilon)}_{ij}e^{(\varepsilon)}_{kl} \cos{(\vec k \cdot \vec r)}
%\eeq
%is the  Hadamard's function of the external gravitational radiation field. That is, when the Hadamard's function of the external gravitational field is zero, the interaction energy will be vanishing at the correspond positions under this order calculations. From Eqs. (\ref{19})-(\ref{20}), we observe that in the near regime, the interaction is always attractive, while in the far regime, it seems that the interaction is also  attractive except for $r=(n+\frac{1}{2})\pi/\omega_R$.  However, when $r\rightarrow(n+\frac{1}{2})\pi/\omega_R$, the next-to-leading term, i.e. the second term in Eq. (\ref{Eab para}) should be taken into account. Therefore, in the far regime, the interaction can also be repulsive  at the vicinity of these positions.
%except that in the far regime, the interaction vanishes when  $r=(n+\frac{1}{2})\pi/\omega_R$, with $n$ being an integer.

Second, if the propagation direction of the  incident external gravitational radiation field is perpendicular to the orientation of the two objects, i.e., $\vec k \cdot \vec r=0$, then Eq. (\ref{fullEab}) yields
\bea\label{Eab per}
\nonumber\Delta E_{AB}=-\frac{I_R}{128\pi r^5}\alpha^{(\varepsilon)}_{A}\alpha^{(\varepsilon)}_{B}
&&\Big[\left(2e^{(\varepsilon)}_{ij}e^{(\varepsilon)}_{ij}-4e^{(\varepsilon)}_{i1}e^{(\varepsilon)}_{i1}+ e^{(\varepsilon)}_{11}e^{(\varepsilon)}_{11}\right) r^4\omega^4_R\cos{\omega_R r}\\
\nonumber &&+2\left(-2e^{(\varepsilon)}_{ij}e^{(\varepsilon)}_{ij}+8e^{(\varepsilon)}_{i1}e^{(\varepsilon)}_{i1} -5e^{(\varepsilon)}_{11} e^{(\varepsilon)}_{11}\right) r^3\omega^3_R\sin{\omega_R r}\\
\nonumber &&+3\left(-2e^{(\varepsilon)}_{ij}e^{(\varepsilon)}_{ij}+12e^{(\varepsilon)}_{i1}e^{(\varepsilon)}_{i1}-15 e^{(\varepsilon)}_{11} e^{(\varepsilon)}_{11}\right) r^2\omega^2_R\cos{\omega_R r}\\
\nonumber &&+3\left(2e^{(\varepsilon)}_{ij}e^{(\varepsilon)}_{ij}-20e^{(\varepsilon)}_{i1}e^{(\varepsilon)}_{i1}+35 e^{(\varepsilon)}_{11} e^{(\varepsilon)}_{11}\right) r\omega_R\sin{\omega_R r}\\
&&+3\left(2e^{(\varepsilon)}_{ij}e^{(\varepsilon)}_{ij}-20e^{(\varepsilon)}_{i1}e^{(\varepsilon)}_{i1}+35 e^{(\varepsilon)}_{11} e^{(\varepsilon)}_{11}\right)\cos{\omega_R r}\Big],
\eea
where we have taken $\hat r_i=(1,0,0)$ and $\vec k =(0,0,k)$.
In the near regime, i.e., $\omega_R r\ll 1$, the leading term of Eq. (\ref{Eab per}) becomes
\beq
\Delta E_{AB}\simeq-\frac{3I_R}{128\pi r^5}\alpha^{(\varepsilon)}_{A}\alpha^{(\varepsilon)}_{B} \left(2e^{(\varepsilon)}_{ij}e^{(\varepsilon)}_{ij}-20e^{(\varepsilon)}_{i1}e^{(\varepsilon)}_{i1}+35 e^{(\varepsilon)}_{11} e^{(\varepsilon)}_{11}\right).
\eeq
So, the interaction energy  decreases as $r^{-5}$ in the near regime. Remarkably, it can be either  attractive or repulsive depending on the polarization of the external gravitational radiation field. For example, the interaction is attractive if the polarization tensor contains only  diagonal elements which may correspond to the $+$ mode of gravitational waves, while it behaves as repulsive when there are only  off-diagonal elements which may correspond to the $\times$ mode. In the far regime, i.e., $\omega_R r\gg 1$, Eq. (\ref{Eab per}) reduces to
\beq
\Delta E_{AB}\simeq-\frac{\omega^4_R I_R}{128\pi r}\alpha^{(\varepsilon)}_{A}\alpha^{(\varepsilon)}_{B}\left(2e^{(\varepsilon)}_{ij}e^{(\varepsilon)}_{ij}-4e^{(\varepsilon)}_{i1}e^{(\varepsilon)}_{i1}+ e^{(\varepsilon)}_{11}e^{(\varepsilon)}_{11}\right)\cos{\omega_R r}.
\eeq
That is, when the propagation direction of external gravitational radiation field is perpendicular to the orientation of the objects, the interaction energy  in the far regime oscillates with a decreasing amplitude which is proportional to $r^{-1}$.  The interaction can  be attractive or repulsive depending on the polarization and frequency of the external gravitational radiation field, and the interobject distance. For a given external field, the interaction periodically behaves between attractive and repulsive as the interobject distance varies.
%Besides, when the frequency of the external field is given, the interaction periodically  behaves  between attractive and repulsive  as the inter-object distance varies.

%{\bf Moreover, in this paper, the gravitational field are required to be weak in the framework of linearized quantum gravity, i.e. $|h_{ij}|\ll 1$. For the external gravitational radiation field, we obtain (in the SI units)
%\beq
%\rho_n\ll\frac{N c^2 \omega_R}{16\pi G \hbar}=\frac{N}{16\pi l_P^2 \lambda},
%\eeq
%where $l_P=\sqrt{\hbar G/c^3}$ is the Planck length, $\lambda=c/\omega_R$ is the wave length of the external gravitational field. In other words, the volume containing the $N$ gravitons should be much larger than $16\pi l_P^2 \lambda$, which can be satisfied in our case.}

\section{Discussion}
\label{sec_disc}
%\setcounter{equation}{0}
%%%%%%%%%%%%%%%%%%%%%%%%%%%%%%%%%%%%%%%%%%%%%%%%%%
In this paper, we investigate the gravitational quadrupole-quadrupole interaction between two gravitationally polarizable objects coupled with a bath of fluctuating gravitational fields in vacuum in the presence of a weak quantized gravitational radiation field, based on the leading order perturbation theory in the framework of  linearized quantum gravity. Our result shows that the interaction energy
%between a pair of ground-state polarizable objects in the weak external gravitational radiation field
behaves as $r^{-5}$ in the near regime and oscillates with a decreasing amplitude proportional to $r^{-1}$ in the far regime. The interaction can be either attractive or repulsive, depending on the polarization, frequency and direction of propagation of the external gravitational field. When the orientation of the two objects is parallel to the propagation direction of the incident gravitational radiation field, the interaction is always attractive in the near  regime,  while in the far regime it can be  attractive or repulsive depending on the frequency of the external gravitational field and the interobject distance.
%except that it vanishes at some specific positions.
When the  orientation of the objects is perpendicular to the propagation direction of the incident gravitational radiation field, the attractive or repulsive property of the interaction depends on the polarization of the incident gravitational radiation in the near regime, while in the far regime it  also depends on the frequency of the external gravitational field and the interobject distance. To conclude, the induced gravitational interaction due to a weak external gravitational field  can be manipulated  by changing the relative orientation of the objects with respect to the propagation direction of the incident gravitational field.

Finally, let us note that there are contributions from other multipole moments to the inter-object interactions (such as monopole-quadrupole cross terms).  In the presence of gravitational waves, a mass monopole oscillates,  and an effective mass quadrupole is formed as seen by a distant observer in analogy to the electromagnetic case \cite{Spruch1994}. Therefore, the monopole-monopole and monopole-quadrupole interactions  due to gravitational vacuum fluctuations and in the presence of external gravitational waves  can also be investigated in the present formalism. We hope to turn to these issues in the  future.

\begin{acknowledgments}

This work was supported in part by the NSFC under Grants No. 11435006, No. 11690034, No. 11805063.

\end{acknowledgments}

\end{document}